# Quantum Annealing for Heisenberg Spin Chains

G.P. Berman[1], V.N. Gorshkov[1,2], and V.I.Tsifrinovich[3]

[1]Theoretical Division, Los Alamos National Laboratory, Los Alamos, NM 87544
[2]Institute of Physics, National Academy of Sciences, Kiev, Ukraine
[3] Physics Department, Polytechnic University, Brooklyn, New York, 11201

Abstract

We suggest using the method of quantum annealing for computing the ground state of the Heisenberg spin chains. Our initial Hamiltonian describes a spin system in a highly non-uniform magnetic field. The initial Hamiltonian gradually transforms into the Heisenberg Hamiltonian, which describes the exchange interaction and the Zeeman interaction with the uniform magnetic field. We demonstrate quantum annealing for 2-, 3-, and 9-spin systems. In particular, we consider the alternate (ferro- and antiferromagnetic) exchange integral. By changing the sign of the exchange integrals we switch from the frustrated to the corresponding non-frustrated spin system. We discuss the similarity and difference between the frustrated and non-frustrated spin systems.

## 1. Introduction

Recently, the quantum annealing method has been widely used to compute the ground state of Ising spin chains [1]. Quantum annealing assumes that the initial Hamiltonian of the system has a simple ground state and does not commute with the Ising Hamiltonian. The initial Hamiltonian smoothly (adiabatically) transforms into the Ising Hamiltonian. In this case a simple initial ground state smoothly transforms into the complicated ground state of the actual Ising Hamiltonian. Typically, the initial Hamiltonian describes the interaction of the spin system with the transverse magnetic field.

The Ising spin chain has numerous applications to mathematically intractable problems. From the other side the Heisenberg spin chains may be used to create atomic-scale magnetic structures, in particular, for the data storage devices. Heisenberg spin chains are actively studied now for applications in nano-technology [2]. Consequently, we consider the application of the quantum annealing to the Heisenberg spin chain. The uniform transverse magnetic field cannot be used now because the isotropic Heisenberg Hamiltonian interacts with any component of the total spin. Thus, we have to consider interaction with non-uniform magnetic field, which does not commute with the Heisenberg Hamiltonian.

## 2. Two-spin Heisenberg system

In order to understand the properties of the quantum annealing computation we start with a trivial two-spin Heisenberg system. We consider the following time-dependent Hamiltonian of the system:

$$\hat{H}(t) = \left[-4J\mathbf{S}_1\mathbf{S}_2 + \gamma\hbar B_0(S_1^z + S_2^z)\right]\frac{t}{\tau} + \gamma\hbar B'(S_1^x - S_2^x)(1-t/\tau), \qquad (1)$$

where $0 < t < \tau$, $J$ is the exchange interaction constant, $\gamma$ is the absolute value of the electron gyromagnetic ratio, $B_0$ is the uniform magnetic field which points in the positive $z$-direction, $B'$ is the magnitude of the non-uniform magnetic field which points in the positive $x$-direction at the location of the first spin and points in the negative $x$-direction at the location of the second spin.

A non-uniform magnetic field is introduced in order to find the ground state of the Heisenberg system using the quantum annealing approach. At time $t = 0$ the Hamiltonian (1) describes two independent spins interacting with the non-uniform magnetic field. Thus, the eigenstate of the Hamiltonian (1) at $t = 0$ represents the first spin pointing in the negative $x$-direction and the second spin pointing in the positive $x$-direction. We choose this state as the initial state for the quantum annealing. At $t = \tau$ the Hamiltonian (1) describes the coupled Heisenberg spins interacting with the uniform magnetic field $B_0$, which points in the positive $z$-direction. If $\tau$ is large enough one can expect that the ground state of the initial Hamiltonian $\hat{H}(t = 0)$ will transfer smoothly to the ground state of the final Hamiltonian $\hat{H}(t = \tau)$. Below we describe our computer simulations for the two-spin system.

We use $B_0$ as the unit of the magnetic field, $\gamma B_0$ as the unit of frequency, $\gamma\hbar B_0$ as the unit of energy, and $1/(\gamma\hbar B_0)$ as the unit of time. In these units we have chosen the following parameters of our Hamiltonian:

$$J = 5, \ B' = 20, \text{ and } \tau = 500. \qquad (2)$$

Positive value of $J$ corresponds to the simplest ferromagnetic ground state. We take the eigenstates of the operators, $S_k^z$, as the basis states and represent the wave function as a superposition of all basis states $|n\rangle$ with the amplitudes $C_n$. The initial values of the amplitudes are

$$C_0 = -½, \qquad C_1 = -½, \qquad C_2 = ½, \qquad C_3 = ½. \qquad (3)$$

Here and below we assume that for a single spin state $|0\rangle$ represents the "down" direction, and state $|1\rangle$ represents the "up" direction. For two or more spins we use the same rules to transfer from the binary to the decimal notation. Thus, for the two spins state $|0\rangle$ represents $|00\rangle$, state $|1\rangle$ represents $|01\rangle$, state $|2\rangle$ represents $|10\rangle$, and state $|3\rangle$ represents $|11\rangle$. Note that we count spins from the left to the right. For example, the state $|01\rangle$ indicates that the first spin points "down" while the second spin points "up".

Fig. 1*a* shows evolution of the probabilities $p_n = |C_n|^2$. One can see that the probability $p_0$ gradually approaches to unity, while all the other probabilities approach to zero. Consequently, the spin system approaches to the ferromagnetic ground state with spin pointing in the negative $z$- direction. Fig. 2*a* shows probabilities $p_n$ for the antiferomagnetic exchange interaction $J = -5$. In this case probabilities $p_1$ and $p_2$ approach the value 0.5 while the two other probabilities vanish. This reflects evolution to the antiferromagnetic state $S = 0$, which is a superposition of the two states $|01\rangle$ and $|10\rangle$ with equal probabilities and phase difference $\pi$. (Phase is not shown in Figs. 1a and 2a.) In Figs. 1*b* and 2*b*, we show also the change of the energy levels for $J = 5$ and $J = -5$, correspondingly. These graphs do not have a direct connection to the quantum annealing. They are presented to better understand the spin systems under consideration.

**3. The Heisenberg system of three and more spins**

For three or more spins we use a Hamiltonian similar to (1)

$$\hat{H}(t) = \left(-4\sum_{k,m} J_{km} \mathbf{S}_k \mathbf{S}_m + \gamma \hbar B_0 \sum_k S_z^k\right)\frac{t}{\tau} + \gamma \hbar B'(1-t/\tau)\sum_k (-1)^{k+1} S_k^x, \qquad (4)$$

At time $t = 0$ the spins point in the direction of their local magnetic field (i.e. in the positive $x$- direction for even $k$, and the negative $x$- direction for odd $k$). We assume a closed chain of spins. If the product of the exchange integrals is an odd number then the spin system is frustrated.

Fig. 3a demonstrates quantum annealing to the ferromagnetic ground state for the 3-spin system with $J_{km}$ = 5. Fig. 3b shows the evolution of the energy levels for the same system. Note that the probability $p_0$ does not increase monotonically as it did for the 2-spin system. Fig. 4*a* shows quantum annealing for the frustrated 3-spin system: $J_{12} = J_{13}$ = 5, $J_{23}$ = -5, and Fig. 4*b* shows the change of the energy levels. In this case, the ground state is a superposition of the states $|1\rangle = |001\rangle$ and $|2\rangle = |010\rangle$ with equal probabilities and phase shift π. One can see that the probabilities $p_1$ and $p_2$ monotonically increase during the process of quantum annealing. Surprisingly, quantum annealing for the frustrated system appears simpler than for the ferromagnetic system.

Next, we consider a more complicated system of 9 spins. The initial energy levels for $t=0$ are shown in Fig. 5. All energy levels except for the lowest and the highest ones are degenerate. Fig. 6 shows the final energy levels at $t=\tau$ for the case of the ferromagnetic exchange interaction, $J_{km}=5$. In the scale of the picture we observe an almost continuous energy spectrum except for the highest levels. The ground state of the system is the ferromagnetic state with the total spin $S=9/2$ and the z-component of the total spin $S^z=-9/2$. The first excited state is the same ferromagnetic state with $S = 9/2$ and $S^z = -8/2$. Correspondingly, the spacing between the ground and the first excited state is exactly one unit, which is not visible in the scale of Fig. 6. Quantum annealing drives the spin system to the ferromagnetic ground state.

Now, we consider a system of 9 spins with alternate exchange integral. This spin system is shown in Fig. 7. The five double lines in the Fig. 7 correspond to positive values of the exchange integral $J=5$, and the four solid lines correspond to negative values, $J=-5$. The final energy spectrum is shown in Fig. 8. The two lowest energy levels have a Zeeman spacing of one unit, the same as for the ferromagnetic case. In the scale of Fig. 8 these two levels are represented by a single dot. These two levels are separated by a gap of approximately 10.8 units from the following quasi-continuous spectrum. The second gap appears near the highest levels. The ground state obtained by the quantum annealing is a complicated superposition of the basis states. (See Fig. 9.) The maximum probability corresponds to the basis state $|102\rangle$: $p_{102} \approx 0.18$. Fig. 10 demonstrates the structure of the state $|102\rangle$: the circles at locations 3,4,7, and 8 show the spins, which point "up".

Finally we discuss the same spin system with the opposite sign of the exchange integrals $J_{km}$. This spin system is frustrated because the number of negative exchange integrals is odd. The energy spectrum of the frustrated spin chain is very similar to that of the non-frustrated one. (See Fig. 11.) At the same time, the structure of the ground state, which is obtained using quantum annealing, is much more complicated. Fig. 12 shows the values of the amplitudes $C_n$. One can see that the ground state of the frustrated spin system has four dominant states instead of a single one for the non-frustrated system. These dominant states are shown in Fig. 13. The corresponding values of the amplitudes $C_n$ are:

$$C_{300} = 0.34, \quad C_{308} = -0.34, \quad C_{306} = 0.32, \quad C_{332} = -0.32. \qquad (5)$$

Thus, we have the two pairs of the dominant states. The two states in a pair have approximately equal probabilities and a phase shift of $\pi$.

## 4. Conclusion

We have demonstrated the successful application of the quantum annealing method to Heisenberg spin chains. In order to use quantum annealing we "apply" a highly non-uniform magnetic field. The corresponding Hamiltonian does not commute with both the Zeeman and the exchange terms of the basic Hamiltonian. First, we demonstrated fast quantum annealing for the simple two- and three-spin systems. Then we considered a more complicated 9-spin system. We have shown that the frustrated spin system has almost the same energy spectrum as the corresponding non-frustrated system but the structure of the ground state for the two systems is significantly different. The non-frustrated system has a single dominant state while the corresponding frustrated system has the two pairs of the dominant states. The states in each pair have approximately equal probability and a phase shift of $\pi$.

**References**

1. Arnab Das, Bikas Chakrabarti (Eds), "Quantum Annealing and Other Optimization Methods", Lecture Notes in Physics, Publisher: Springer Berlin / Heidelberg, V. 679/ 2005.
2. Mark Wilson, Physics Today, **59,** (2006) 13.

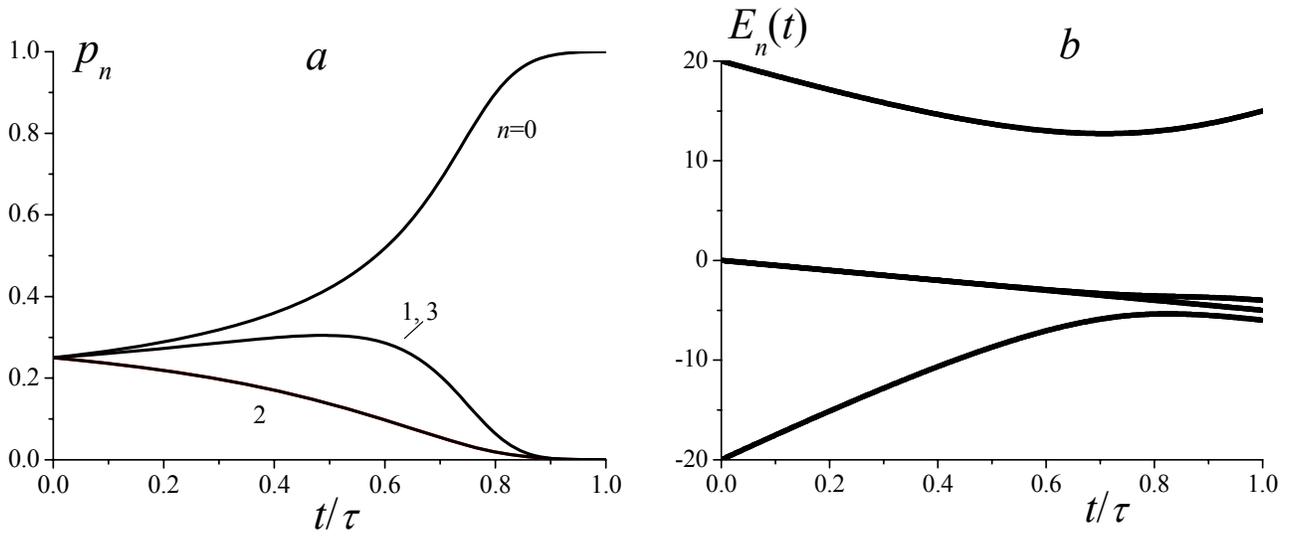

Fig. 1. *a*-probabilities of the basis states, which contribute to the ground state of the 2-spin system with the ferromagnetic exchange interaction, as a function of time in the process of the quantum annealing; *b*- the energy levels of the spin system as a function of time.

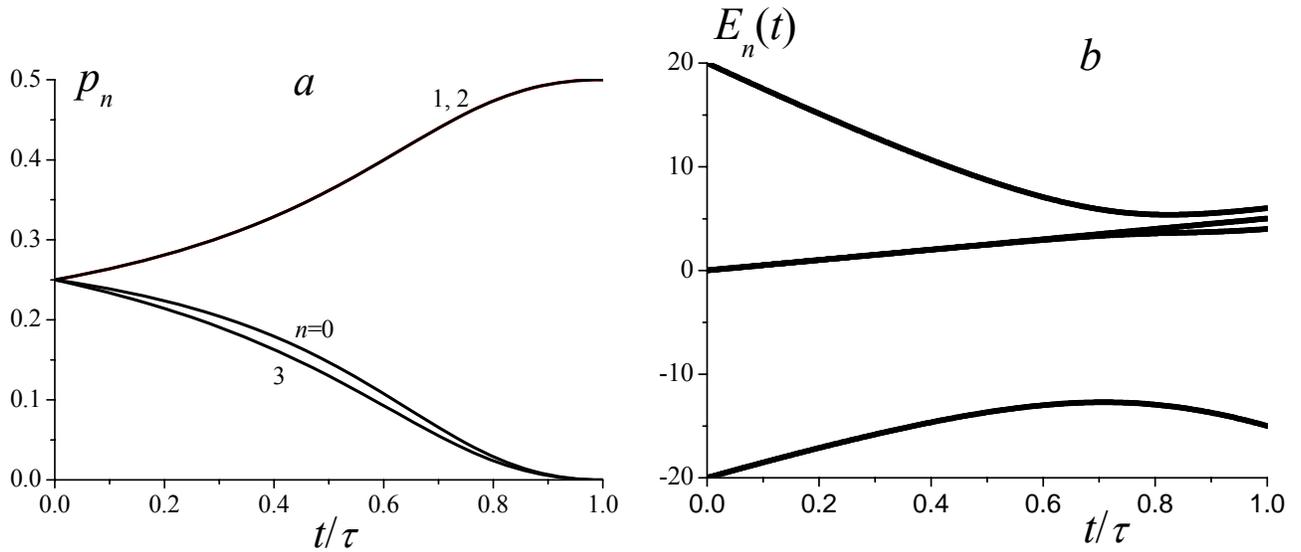

Fig. 2. The same as in Fig. 1 for the 2-spin system with the antiferromagnetic exchange interaction.

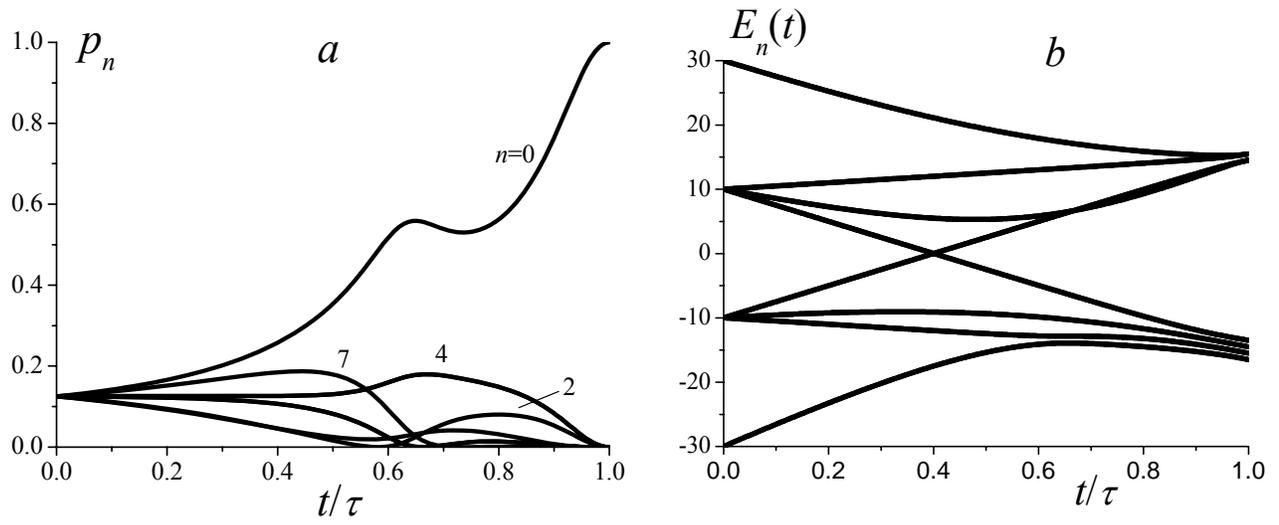

Fig. 3. The same as in Fig. 1-2 for the three spin system with the ferromagnetic exchange interaction.

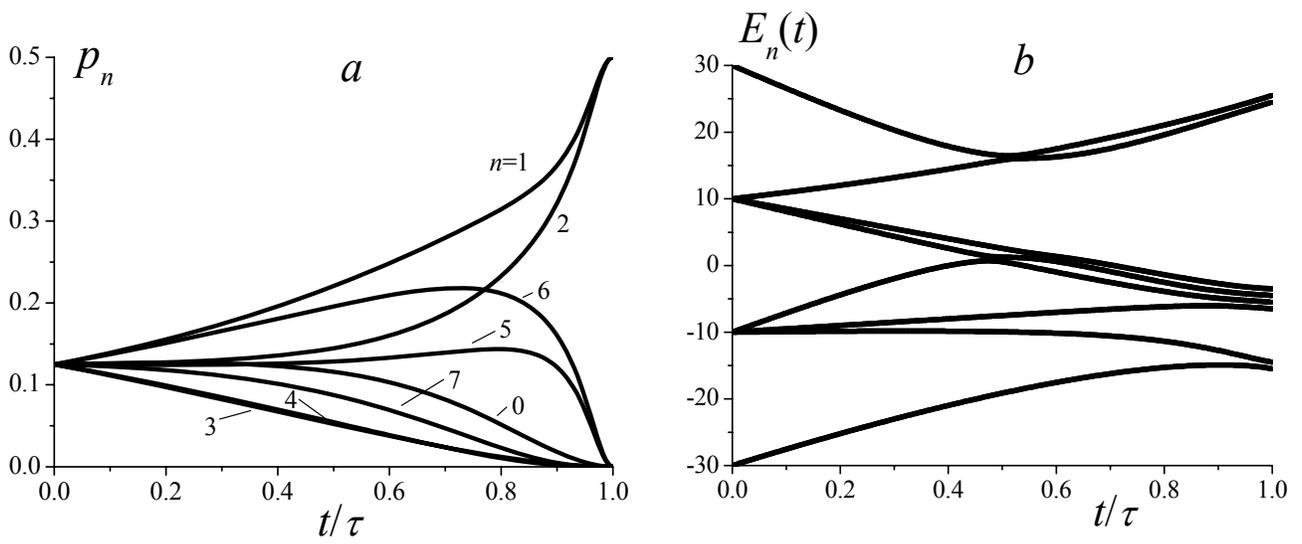

Fig. 4. The same as in Figs. 1-3 for the frustrated 3-spin system.

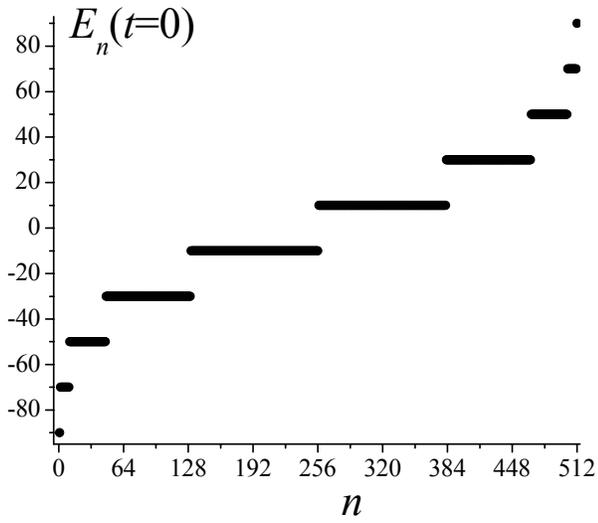

Fig. 5. The energy levels of the 9-spin system at the beginning of the quantum annealing.

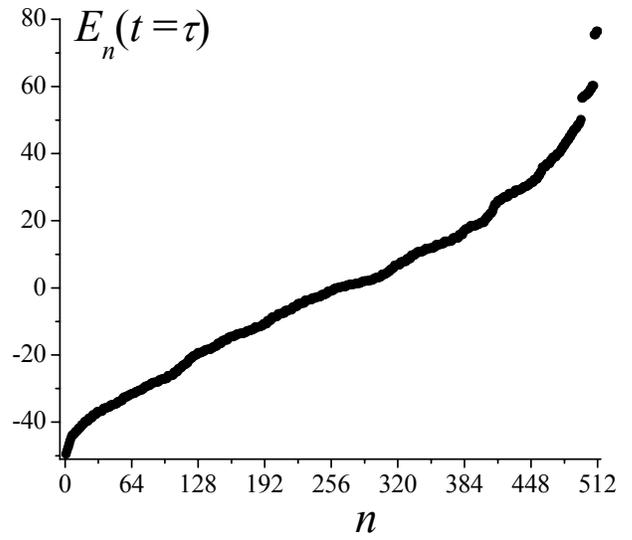

Fig. 6. The energy of the stationary states of the 9-spin system with the ferromagnetic exchange interaction

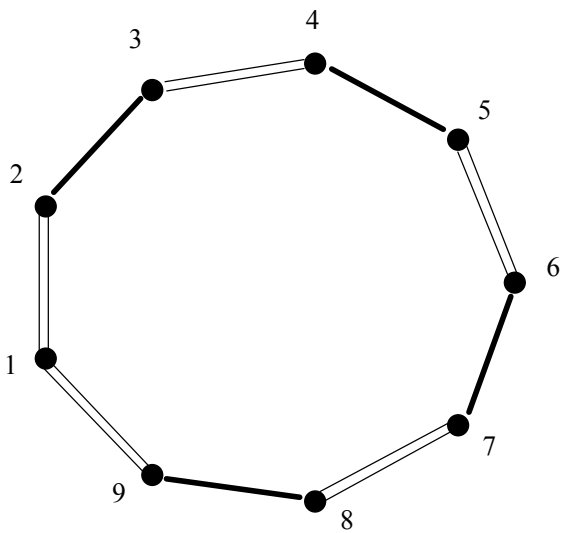

Fig. 7. The 9-spin chain with the alternate ferro- antiferromagnetic coupling (solid line - $J = -5$, double line - $J = 5$).

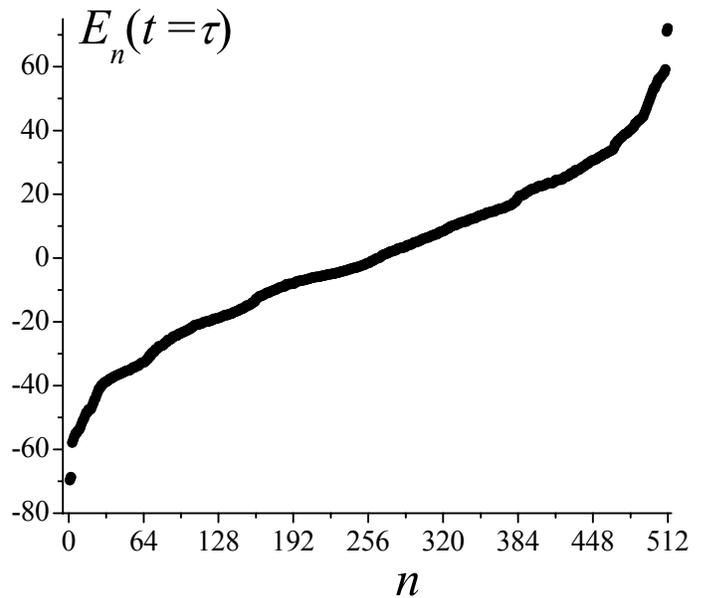

Fig. 8. The energy of the stationary states of the non-frustrated 9-spin system with the alternate exchange interaction.

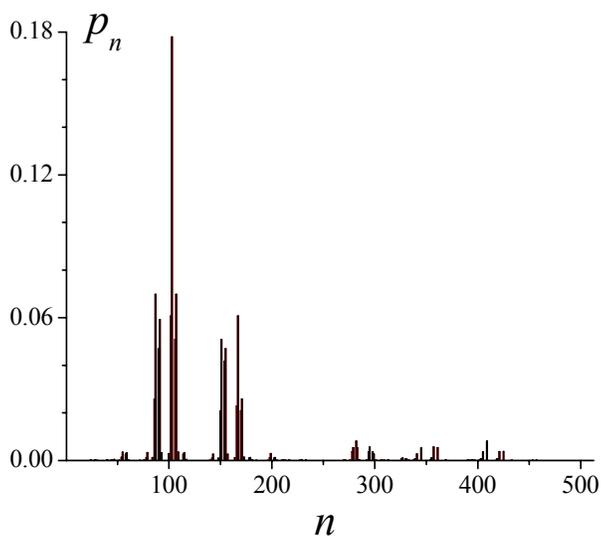

Fig.9. Probabilities of the basis states, which contribute to the ground state of the non-frustrated 9-spin system with the alternate exchange interaction.

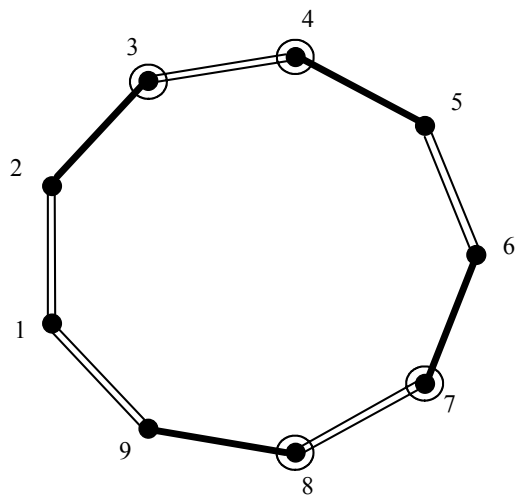

Fig.10. The structure of the basis state $|102\rangle = |001100110\rangle$, which makes the main contribution to the ground state. Circles indicate spins pointing "up".

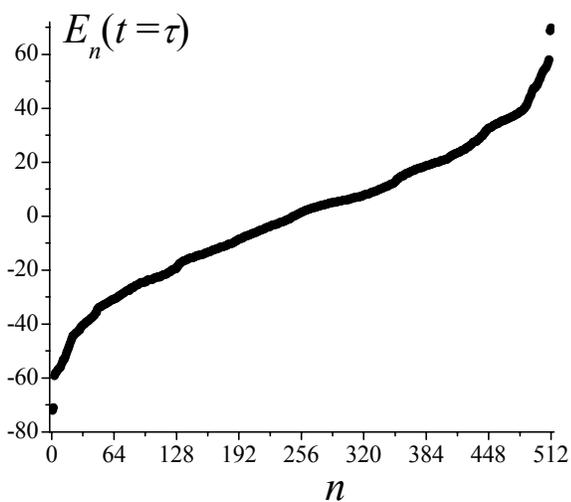

Fig. 11. The energy of the stationary states of the frustrated 9-spin system with the alternate exchange interaction.

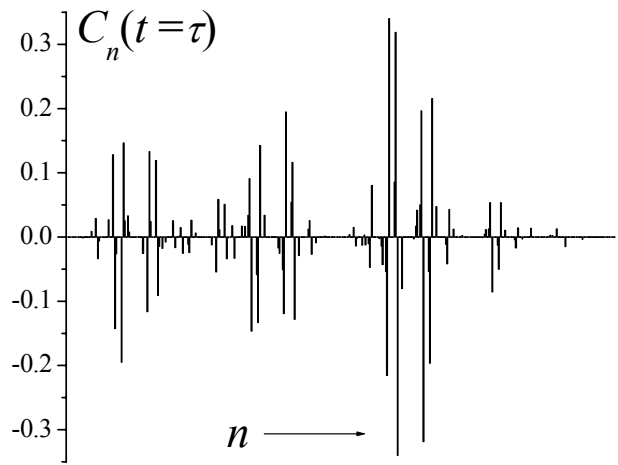

Fig. 12. Amplitudes of the basis states, which contribute to the ground state of the frustrated 9-spin system with the alternate exchange interaction.

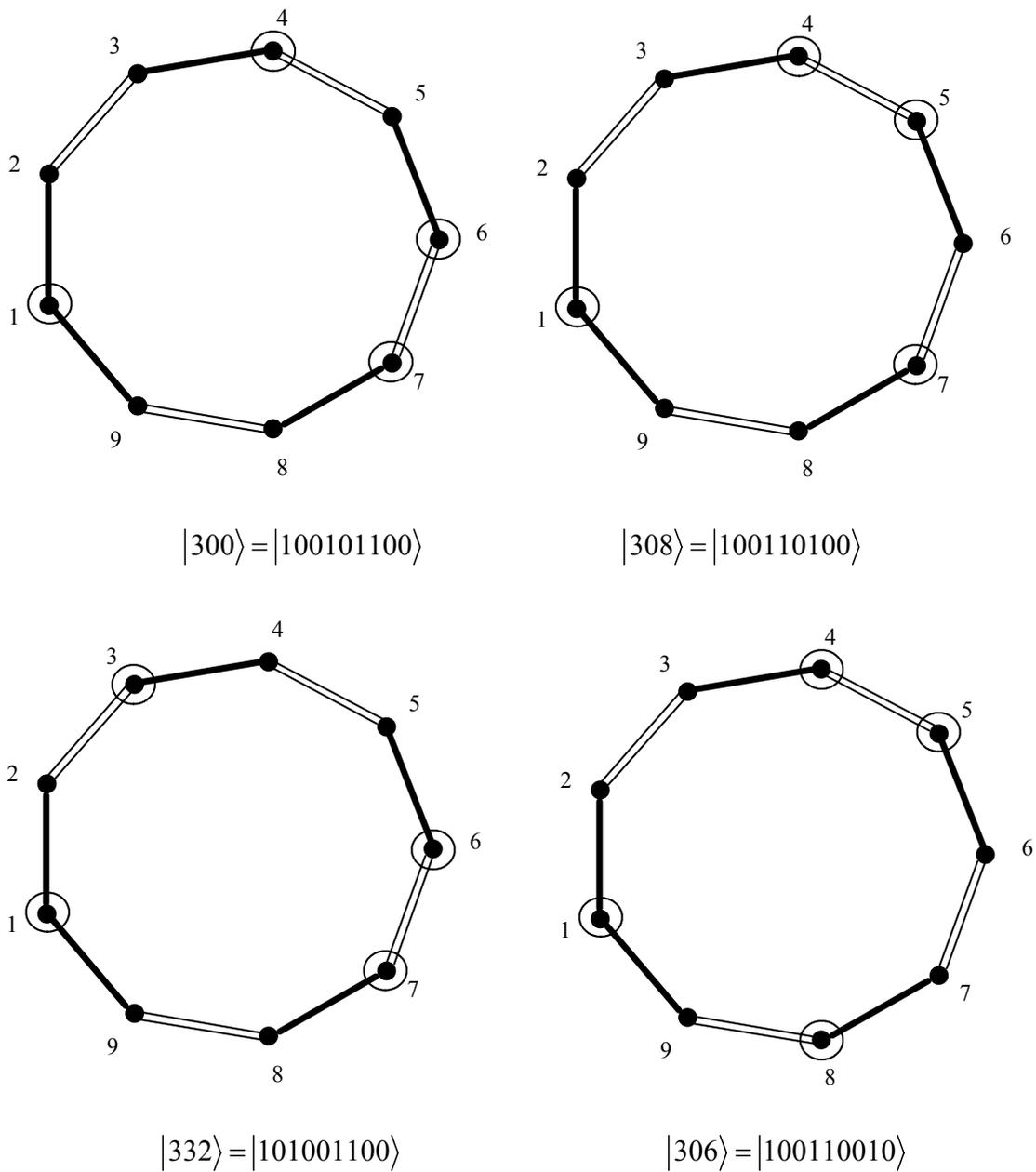

Fig. 13. The structure of the basis states, which make the main contribution to the ground state. Circles indicate spins pointing "up".